\documentclass[11pt]{ar}
\usepackage[english]{babel}
\usepackage{psfig,graphicx}
\usepackage{longtable}

\onecolumn
\begin{document}

\title{High resolution optical spectroscopy of an LBV-candidate inside the Cyg\,OB2
            association}

\author{V.G.~Klochkova,\thanks{\email{valenta@sao.ru}}
 \and E.L.~Chentsov}

\date{\today}	     

\institute{Special Astrophysical Observatory RAS, Nizhnij Arkhyz,  369167 Russia}

\abstract{For the first time, we obtained the high-resolution ($R=15000$
and 60000) optical spectra for the extremely luminous star No.\,12,
associated with the IR--source IRAS\,20308+4104, a member of the Cyg\,OB2
association. We have identified about 200 spectral features in range
4552--7939\,\AA\AA, including the interstellar NaI, KI lines and
numerous DIBs, which are the strongest absorption lines in the spectrum,
along with the HeI, CII, and SiII lines. A two-dimensional spectral
classification indicates that the star's spectral type is $B5\pm 0.5 Ia^+$.
Our analysis of the $V_r$ data shows the presence of a $V_r$
gradient in the stellar atmosphere, caused by the infall of matter onto
the star. The strong H$\alpha$ emission  displays broad Thompson wings and
time-variable core absorption, providing evidence that the stellar wind is
inhomogeneous, and a slightly blue-shifted P~Cyg type absorption profile.
We concluded that the wind is variable in time.
}

\authorrunning{Klochkova \& Chentsov}
\titlerunning{Spectroscopy of Cyg\,OB2-No.\,12}

\maketitle

\section{Introduction}

Study of mass loss and chemical composition variations in the surface
layers are important for our understanding of the evolution of massive
stars. It is crucial to establish the evolutionary stage and the star's
luminosity. One widely used approach to stellar evolution study is to
investigate stars inside clusters and groups: the evolutionary stage, age,
and luminosity can be determined more reliably for group members, whereas
these characteristics are rather uncertain for field stars. It is
especially important to study group members that are rare, such as
LBV--stars or Wolf--Rayet stars. From this point of view, the Cyg\,OB2 (or
VI~Cyg) association, with an age of several million years, is of special
interest. The Cyg\,OB2 association contains a group of high luminosity
stars with very high masses (about 100 ${\mathcal M_{\odot}}$), and that
is so large that the association may be considered as a young globular
cluster [\cite{Comeron}]. According to Massey et al. [\cite{Massey}],
unevolved $O$-- and $Of$--stars have been identified in the association,
as well as an LBV candidate--the variable star No.\,12 from the list by
Schulte [\cite{Schulte}]. Hereafter we will refer to this star as
Cyg\,OB--No.\,12. The star Cyg OB2--No.\,12 is among the brightest OB
stars in the IR, due to the presence of circumstellar matter lost by the
star via its strong wind [\cite{Wendker}]. The star is identified with the
IR source IRC+40430\,=\,IRAS\,20308+4104. For a association distance of $m
- M = 11\lefteqn{.}^m2$, the star luminosity is $log(L/L_{\odot}) = 6.26$
[\cite{Jager}] and its bolometric absolute magnitude is $M_{bol} = -11^m$
[\cite{Massey2}]. Massey and Thompson [\cite{Massey2}] classified the star
as B5\,Ie, and Souza and Lutz [\cite{Souza}] as B8\,Ia, whereas the
authors [\cite{Humphreys}] considered it as one of brightest
A--supergiants in Galaxy. Later, Massey et al. [\cite{Massey}] confirmed
the LBV candidate status of Cyg\,OB2--No.\,12 based on its membership in
an association whose turnoff point is near that for the Milky Way's most
massive stars. The luminosity of Cyg\,OB2--No.\,12 from its association
membership indicates that the star is one of the four most luminous stars
in our Galaxy (see, for instance, the diagram for S\,Dor stars in
\cite{Genderen} and in the review of de~Jager [\cite{Jager}]). However,
the star's visible flux is strongly attenuated due to the large distance
to the association and the presence of strong extinction: the observed
magnitudes are $B = 14\lefteqn{.}^m41$, $V = 11\lefteqn{.}^m40$. It was
recognized long ago that, apart from being distinguished by its high
luminosity, Cyg OB2--No. 12 also displayed very high reddening
[\cite{Morgan, Sharpless}]. Based on their photometric study of a stellar
sample in Cyg\,OB2, Torres-Dodgen et al. [\cite{Torres}] estimated the
distance $m - M = 11\lefteqn{.}^m2 \pm 0\lefteqn{.}^m2$ and its distance
$d = 1.7 \pm 0\lefteqn{.}^m3$\,kpc, and confirmed that the high
interstellar reddening for the association members (with a mean value $E(B
- V)=1\lefteqn{.}^m82$) was satisfied by a normal law. The exception is
Cyg\,OB2--No.\,12, whose visual extinction exceeds $10^m$\,(!). Obviously,
spectroscopic studies of stars experiencing such a high degree of
reddening are possible only thanks to the relatively small distance to the
association and the high absolute luminosities of many of its members.

Lozinskaya et al. [\cite{Lozin}] also emphasized that the group of massive
stars inside the Cyg\,OB2 association probably possessed the Galaxy's
strongest stellar winds, which are capable of significantly affecting the
ambient interstellar gas over some two to three million years.
Cyg\,OB2--No.\,12 is one of the few late B stars known to radiate thermal
radio emission [\cite{Wendker}]. The variability of its radio flux is
surprising [\cite{Bieging}]. White and Becker [\cite{White}] estimated the
mass loss rate to be $4 \times 10^{-5}\mathcal{M}_{\odot}$/yr, that is too
high for a normal supergiant, but consistent with the extreme luminosity
of Cyg\,OB2--No.\,12. According to the criterion suggested by Humphreys and
Davidson [\cite{Humphreys}], a mass loss rate that high indicates that
Cyg\,OB2--No.\,12 is an LBV star. Through his modeling of the IR spectral
energy distribution, Leitherer et al. [\cite{Leitherer}] estimated the
star's effective temperature to be $T_{eff}$=13600\,K and the envelope's
electron temperature to be $T_e$=5000\,K. The combination of a hot
atmosphere and a cool, dense envelope found for Cyg\,OB2--No.\,12 is not
unique: such structures are known for S\,Dor stars [\cite{Schulz}]. The
light from Cyg\,OB2--No.\,12 is polarized [\cite{Nord}]. The
broadband polarimetry of Schulz and Lenzen [\cite{Schulz}] at
0.3--1.1\,$\mu$ displayed
linear polarization exceeding 10\%, providing evidence for a nonspherical
distribution of the circumstellar material, and hence for a nonspherically
symmetrical stellar wind. The accumulated observations show that
Cyg\,OB2--No.\,12 is a crucial object for studies of late evolutionary
stages for massive stars, creating the need for high-resolution optical
spectroscopy that would make it possible to classify the spectrum, and
thereby refine estimates of the star's fundamental parameters and the
characteristics of its stellar wind.

\section{Observations and data reducing} 

Our spectroscopic observations of Cyg\,OB2--No.\,12 were acquired with the
6\,m telescope of the Special Astrophysical Observatory (Russian Academy of
Sciences) using echelle spectrographs. Our first set of observations,
obtained on June 12, 2001 using the PFES spectrometer [\cite{Panchuk}]
with a 1040x1170-pixel CCD at the primary focus, yielded a spectrum in the
interval 4542--7939\,\AA\AA{} with a resolution of $R =\lambda/\Delta\lambda
\approx 15000$ (20\,km/s). A second observing run  was made on
April 12, 2003 at the Nasmyth focus using the NES spectrograph
[\cite{Panch}] equipped with an image slicer [\cite{Panchuk3}]. We
obtained a spectrum in the range 5273--6764\,\AA\AA{} with a resolution of
$R \approx 60000$ (5\,km/s) using a 2048x2048  CCD. We used the PFES
spectrometer to acquire spectra of bright, luminous B stars for use in
spectral classification (see Section \ref{Sp} for details). We removed
cosmic-ray traces via a median averaging of two consecutive exposures. The
wavelength calibration was performed using the spectrum of a ThAr
hollow-cathode lamp. We recorded the spectrum of the hot, rapidly rotating
star HR\,4687, which has broad lines, each night to subtract the telluric
spectrum. The preliminary reduction of the CCD images of our echelle
spectra (removal of cosmic rays, background subtraction, wavelength
calibration, and extraction of the spectral orders) was performed using
the MIDAS (98NOV version) ECHELLE package. The final reduction (continuum
normalization, measurements of radial velocities and equivalent widths for
various spectral features) was done using the DECH20 program package
[\cite{Galaz}].

\section{Results} 

\subsection{Main peculiarities of the  Cyg\,OB2--No.\,12 spectrum}\label{Sp}

The main features of Cyg\,OB2--No.\,12 are evident even in low-resolution
spectra [\cite{Morgan}]: it is an early-type, very luminous star with very
strong H$\alpha$ emission and very strong diffuse interstellar bands
(DIBs). The Na\,D1,2 doublet lines display no obvious peculiarities.
Wendker and Altenhoff [\cite{Wendker}] note that the H$\alpha$ profile is
probably variable. Our high-resolution spectra permits us for the first
time to make detailed line identifications, classify the spectrum, and
measure radial velocities. Table\,1 presents identifications of the
spectral features in the spectra obtained on June 12, 2001 and April 12,
2003, along with their equivalent widths $W$, residual intensities $r$,
and heliocentric radial velocities $V_r$. Figs.\,1 and 2 display fragments
of the spectra of Cyg\,OB2--No.\,12 and of the hypergiant HD\,168625
(B6\,Ia$^+$, $M_v = -8\lefteqn{.}^m5$). The spectrum of HD\,168625 was
taken on June 19, 2001 with the PFES spectrometer. These spectra are very
similar, demonstrating that the stars have similar temperatures and
luminosities. In addition, DIBs are equally well represented in these
spectra and have similar intensities. The strongest of them, the 5780 and
5797\,\AA{} bands, dominate in Fig.\,1, but even the weak DIBs at 5766,
5773\,\AA{}, etc. have comparable intensities to the stellar NII, AlIII,
and SiIII absorption. The DIBs at 6376 and 6379\,\AA{} in Fig.\,2 are
almost as deep as the SiII\,(2) absorption. The majority of spectral
features are shallow absorption lines with depths of 0.02-0.03 of the
continuum level, whose depths, equivalent widths, and radial velocities
are uncertain (marked with colons in Table 1). The signal-to-noise ratio
in the blue was not high, and we used nonstandard criteria for our
quantitative spectral classifications. The sets of classification criteria
and the wavelength intervals used for 2001 and 2003 overlap only
partially. The classification techniques are described in more detail in
[\cite{Prep183}]. Here we note only that the HeI, CII, NII, AlIII, SiII, SII,
and FeII lines were used. The $W(Sp)$ calibration relations were based on
the supergiant standard stars $\epsilon$\,Ori (O9.5Ia), HD\,13854
(B1.2Iab), HD\,14134 (B2.2Ia), HD\,206165=9\,Cep (B2.5Ib),
HD\,198478=55\,Cyg (B3.4Ia), HD\,164353=67\,Oph (B3.6Ib, $M_v =
-5\lefteqn{.}^m6$), HD\,58350=$\eta$\,CMa (B3.9Ia), HD\,13267=5\,Per
(B5Ia), HD\,15497 (B6Ia, $M_v = -7\lefteqn{.}^m0$), HD\,183143 (B7.7Ib,
$M_v = -8^m$), HD\,34085=$\beta$\,Ori (B8.2Ia, $M_v = -7\lefteqn{.}^m3$),
and HD\,21291 (B9.3Ib, $M_v = -6\lefteqn{.}^m9$). A comparison of the
depths and equivalent widths in our two spectra shows that,
although the depths were slightly smaller in 2001 than in 2003 (due to the
lower resolution), the equivalent widths did not differ systematically. It
is thus natural that the spectral types derived for Cyg\,OB2--No.\,12 for
the 2001 and 2003 data were the same within the errors: B5.0$\pm$0.5 and
B4.8$\pm$0.5, respectively. The luminosity class for both Cyg\,OB2--No.\,12
and HD\,168625 is Ia$^+$. Evidence for the star's high luminosity is also
provided by the high intensity of the OI\,7773\,\AA{} IR--triplet, whose
equivalent width is $W_{\lambda}$=1.14\,\AA{}, corresponding to an absolute magnitude
of $M_v < -8^m$.

\subsection{The H$\alpha$ line profile}

Based on two spectra with 3\,\AA{} resolution taken three days apart, Souza
and Lutz [\cite{Souza}] concluded that the radial velocity derived from
the H$\alpha$ emission line was variable. An absorption line at 6532\,\AA{}
was detected in one of the spectra, whose posi- tion corresponds to an
expansion velocity of about $-(1400 \div 1500)$\,km/s. The spectrum with
7\,\AA{} resolution presented in [\cite{Nord}] shows a similar absorption
feature (6526\,\AA{}). Our high-resolution spectra enabled us to study the
fine structure of the H$\alpha$ emission (Fig.\,3). The profile has broad
wings extending to at least $\pm$1000\,km/s is slightly asymmetric, with a
weak absorption visible in its left wing, suggestive of a P\,Cygni type
profile; and displays absorption features with variable shape in the core.
These absorption features correspond to transitions in the H$\alpha$ line
rather than to the telluric spectrum, whose contribution was carefully
removed from the spectrum. We are inclined to interpret the emission wings
of the H$\alpha$ profile as being due to Thompson scattering on the
envelope's free electrons [\cite{Marl, Castor, Bernat}]. Stark emission
wings are formed in denser media, such as the deexcitation region behind
the shock front in the atmosphere of W\,Vir [\cite{Lebre}]. Following Wolf
et al. [\cite{Wolf}], let us estimate the efficiency of Thompson
scattering for a hot envelope in which the hydrogen is completely ionized.
In accordance with [\cite{White}], let us assume a mass-loss rate of
$4\times10^{-5}{\mathcal M}_{\odot}/$yr and a radius for the envelope
equal to twice the stellar radius ($R_{\star} = 338 R_{\odot}$
[\cite{Bieging}]). It is more difficult to estimate the escape velocity
based on the two absorption features observed in the H$\alpha$ wing
[\cite{Souza, Nord}]. The value of 1400\,km/s used in [\cite{White}] seems
too high. Note that the shell forming the narrow absorption feature must
be spatially separated from the shell forming the P\,Cygni profile. It is
not clear why the presence of this shell is not manifest in the spectrum
taken three days later [\cite{Souza}]. The distance covered at such a
speed over three days is about $3.7 \times 10^{13}$\,cm, which is a factor
of four higher than the envelope's radius estimated in an adiabatic
cooling wind model, based on observations of free--free emission
transitions at 6\,cm [\cite{White}]. This would imply that the optical
spectral features are formed further from the star than the radio
emission. This discrepancy can be removed if we reject the assumption
about spherical symmetry for the envelope. In fact, if the wind extends
differently in different directions, its temperature in a spherical
approximation will be lower than the real temperature, as is observed
(5000$\pm$ 1500\,K [\cite{White}]). For the above high escape velocity and
large radius, the optical depth to Thompson scattering in the spherically
symmetric approximation [\cite{Wolf}] is insignificant, about 0.01. If the
escape velocity is lower by an order of magnitude, as is suggested in
[\cite{Bieging}] (see also below our estimate of the highest wind
velocity), and the wind is not spherically symmetric, the Thompson
scattering optical depth increases by more than an order of magnitude. We
conclude that the absorption features we have detected at the peak and in
the blue wing of the H$\alpha$ line are variable, even after we have taken
into account the different spectral resolutions of our observations.

\subsection{Interstellar bands}

The strength of the interstellar (and circumstellar) extinction of the
light from Cyg\,OB2--No.\,12 has made the star a popular target for
studies of interstellar spectral features. In particular, the strong
absorption towards Cyg\,OB2--No.\,12 is of considerable interest from the
point of view to investigate the discrete distribution of the absorbing
material (cf., for example, Scappini et al. [\cite{Scappini}] and
references therein). For the first time Souza and Lutz [\cite{SouzaLutz}]
detected IR--bands of the $C_2$ molecule, and molecular features in the
spectrum of Cyg\,OB2--No.\,12 were subsequently analyzed in a number of
studies. In their high-spectral resolution study, Gredel and M\"unch
[\cite{Gredel}] detected a four-components structure in the IR (1,0) band
of the $C_2$ molecule's Phillips system, which displayed a range of
velocities from $-10.6$ to +13.2\,km/s, whereas Chaffee and White
[\cite{Chaffee}] observed two components in the KI line with velocities
from $-12.9$ to $-4.3$\,km/s. Based on echelle spectra, Gredel et al.
[\cite{Gredel2}] analyzed the physical conditions for the formation of the
$C_2$ and $CN$ molecules in detail, and also detected an interstellar RbI
line. We also note that the star's spectrum is densely populated by
interstellar features, as can clearly be seen in Figs.\,1 and 2. As follows
from Figs.\,1 and 2 and Table\,1, the intensities of the main DIBs and of
the NaI lines are higher than those of the strongest photospheric
absorption lines (HeI, CII, SiII) in the spectrum of Cyg\,OB2--No.\,12. We
discuss the structure of the NaID lines below, in connection with the
systemic velocity of Cyg\,OB2--No.\,12.

\subsection{The radial velocities pattern}

Table\,2 presents radial velocities for individual lines and averages for
groups combining lines with similar residual intensities. Possible
systematic errors for the velocities in Tables\,1 and 2 estimated from the
telluric and interstellar lines are within 2 and 1\,km/s for the June 12,
2001 and April 12, 2003 spectra, respectively. It is also necessary to
take into account the differing spectral resolutions when comparing
$V_r$\,(NaI) for the 2001 and 2003 spectra. Fig.\,4 shows that, in the
April 2003 spectrum, the main component with $r \approx 0.01$ and $V_r =
-9$\,km/s is clearly separated from a component with $V_r = -34$\,km/s
that is half as strong. The June 2001 PFES spectrum does not separate
these components, but the asymmetry of the profile is appreciable, with
the blue wing being less steep. The random measurement errors for
individual lines can be judged from the scatter of the circles in the
upper panels of Fig.\,5, with the following caveats. The velocities were
measured from absorption cores, i.e., from the lowest parts of the
profiles, with exclusion of the very deepest portions having much lower
intensity gradients. Of the two lines whose profiles are displayed in the
bottom panels of Fig.\,5, the core in the June 2001 spectrum is sharper
for the HeI absorption line, so that its position can be measured more
accurately, whereas the SiII absorption line is sharper in the April 2003
spectrum. The SiII (2) lines (open circles in Fig.\,5) are obviously blue
shifted relative to other absorption lines with the same depths, but the
other lines may also possess small mutual shifts, increasing the scatter
of the data points in the $V_r(r)$ diagrams. In addition, many of the
lines are asymmetric (as is also clearly visible in Fig. 5), and a small
shift in $r$ has a considerable influence on the resulting value of $V_r$.

\subsubsection{\small The systemic velocity}

Unfortunately, there are no data on the radial velocities of any stars of
the association besides Cyg\,OB2--No.\,12 itself and the spectroscopic
binary Cyg\,OB2--No.\,5, whose velocity is very uncertain [\cite{Bohan}].
However, a rough estimate of the systemic velocity, $V_{sys}$, for
Cyg\,OB2--No.\,12 is possible based on the differential rotation of the
Galaxy. In our case, this estimate is facilitated by the fact that the
dependence of the velocity on distance is weak in the direction toward
Cyg\,OB2 (along the Cygnus arm), and almost disappears in the region of the
Cyg\,OB2 association [\cite{Brand}]. Under this circumstance we used a
wider range of acceptable distances for stars, HII regions, and cool
gaseous clouds whose radial velocities are to be used to estimate $V_{sys}$.
The relation between the heliocentric radial velocity, $V_r$, and Galactic
longitude, $l$, for stars between 1 and 2\,kpc selected from the catalogs
[\cite{Hump, BS1, BS2}] gives the mean value $V_r = -12 \pm 3$\,km/s for the
longitude of Cyg\,OB2 ($l=80{\degr}$). The peak of the interstellar H$\alpha$
emission profile at $l=80{\degr}$ is also at $V_r = -12$\,km/s [\cite{Lozin,
Reyn}], and the highest intensity of the $CO$ radio emission is at $V_r =
-9$\,km/s [\cite{Dame}]. A similar $V_r$ value for the Cyg\,OB2
association (about $-10$\,km/s) results from the detailed analysis of the
motions of stars and interstellar matter in the Cygnus arm carried out by
Sitnik et al. [\cite{Sitnik}].

Let us supplement the above $V_r$ estimates with those obtained
specifically for Cyg\,OB2--No.\,12. According to McCall et al.
[\cite{McCall}], the $CO$ emission has two intensity peaks, with $V_r =
-7$ and $-2$\,km/s, and the interstellar absorption profiles of KI, $CO$,
$C_2$, etc. have up to five components with $V_r$ values from $-13$ to
12\,km/s. The profiles of the interstellar NaI doublet in our high
resolution spectrum (Fig.\,4) have two main components: the strongest is
saturated and has $V_r =-9$\,km/s, while the shallower one is blueshifted
and has $V_r = -34$\,km/s. A similar division into components with the
same intensity ratio and velocities ($-11$ and $-35$\,km/s) is observed in
the NaI\,D2 line for the nearby hypergiant P\,Cygni [\cite{Hobbs}]. The
blue shifted component is probably circumstellar, and is associated with
the stellar wind. Judging from its velocity, the main absorption feature
is formed immediately in front of the star, in the gas complex that Sitnik
et al. [\cite{Sitnik}] call BB. There are dips at the bottom of this
absorption feature, whose velocities ($-13$ and $-6$\,km/s) coincide with
those for the components of the KI line according to Chaffee and White
[\cite{Chaffee}] and McCall et al. [\cite{McCall}]. Finally, we should
include our velocity estimates for stationary envelope emission lines (the
first rows of Table\,2). The weak FeII~7513\,\AA{} emission line, which
is unfortunately the only one, gives $V_r \approx -12$\,km/s; the upper, symmetric
part of the H$\alpha$ profile, which experiences minimal distortion from
the absorption components, gives $V_r = -10$, . . . $-15$\,km/s.

As expected, all these estimates are close to each other. We adopt
$V_{sys}=-11 \pm 2$\,km/s as the systemic velocity for Cyg\,OB2--No.\,12.

\subsubsection{\small Temporal and line-to-line variations of $V_r$}

Fig.\,5 and the data in Table\,1 show that the radial velocities measured
from the absorption cores vary in time and change with the line
intensity. In both our spectra, the weakest lines ($r\rightarrow 1$) yield
$V_r$ values lower than $V_{sys}$ (by 5 and 14\,km/s, respectively, in
2001 and in 2003), testifying to variable rates of expansion of the layers
where these lines are formed. Stronger lines show positive shifts relative
to weaker ones, with the exception of the SiII lines, marked in Fig.\,5 as
open circles. The slope of the filled-circle chain in the left panel of
Fig.\,5, corresponding to June 12, 2001, is quite evident, whereas it is
poorly represented in the upper right panel of Fig.\,5 due to the limited
spectral range and some deficiencies of the April 12, 2003 data. The core
of the deepest available absorption line, HeI~5876\,\AA{}, shows the
highest $V_r$, which exceeded $V_{sys}$ by 10\,km/s on June 12, 2001;
according to McCall et al. [\cite{Oka}], this shift reached values as
large as 22\,km/s.

\subsubsection{\small Manifestations of the wind}

The stellar wind from Cyg\,OB2--No.\,12 is manifest itself most clearly in
the H$\alpha$ profile. Fig.\,3 shows that the profile shape varies with
time, but its principal features are preserved: strong emission, with a
dip at the short-wavelength slope, a sheared peak, and extended Thompson
wings. The blueshifted absorption is barely visible in the June 2001
spectrum and is more pronounced in the April 2003 spectrum, but can be
traced at least to $V_r = -160$\,km/s in both cases; i.e., to the same
limit that is reached by the blue wings of the absorption lines presented
in Fig.\,5. The wind's velocity limit is about 150\,km/s. The intensity
inversions in the upper part of the H$\alpha$ profile are especially
interesting. They indicate that the wind from Cyg\,OB2--No.\,12 is not
uniform: in addition to the high velocity material noted above, it
contains a fair amount of material that is nearly stationary relative to
the star, or is even falling onto the stellar surface. The coexistence of
lines with direct and inverse P\,Cygni profiles in the same spectrum, and
even combinations of such features in the profile of the same line, has
been noted for some LBVs at their maximum brightness [\cite{WStahl,
Stahl}]. Such a behavior leads us to reject the spherically symmetric wind
model. It is possible that the slow part of the wind also contributes to
the absorption profiles; spectroscopic monitoring is needed to check this
hypothesis. So far, this possibility is supported by the coincident
velocities for the central dips of the H$\alpha$ line and the wellformed cores of
strong absorption lines (HeI~5876\,\AA{} in 2001 and SiII~6347\,\AA{} in 2003),
as well as by the fact that the shift of all the absorption lines in the
2003 spectrum towards short wavelengths relative to their positions in
2001 was accompanied by a similar shift of the central dip in H$\alpha$. At any
rate, both the hydrogen lines and the strongest absorption lines in the
visual spectrum Cyg\,OB2--No.\,12 (for the available part of the spectrum,
these are the HeI and SiII lines) are partially formed in the wind.

\section{Conclusions}

Our detailed spectroscopy of the extremely luminous star Cyg\,OB2--No.\,12
in the stellar association Cyg\,OB2 at 4552--7939\,\AA\AA{} has enabled us
to identify about 200 spectral features, including numerous interstellar
features (NaI, KI, and DIBs). Using spectral classification criteria
developed for red wavelengths, we have determined the star's spectral type
($B5\pm0.5$) and luminosity class ($Ia^+$). The intensity of the IR oxygen
triplet, OI~7773\,\AA{}, leads to an absolute magnitude for the star $M_v
< -8^m$. Our analysis of the radial-velocity pattern indicates the
presence of a radial velocity gradient in the atmosphere due to the infall
of matter onto the star. We have detected fine structure of the profile of
the strong H$\alpha$ emission: broad emission wings (to $\approx
1000$\,km/s), absorption at the peak that varies with time, and weak
P\,Cygni absorption features corre sponding to expansion velocities up to
150\,km/s. The intensity inversion at the H$\alpha$ peak provides evidence
that the stellar wind is not uniform: in addition to that part of the wind
that moves away from the star, there is also material that is at rest
relative to the star, or is even falling onto the stellar surface. The
radial velocities measured from the absorptionline cores vary with time
and with the line intensity. The weakest lines ($r\rightarrow 1$) in our
two spectra give $V_r$ values lower than $V_{sys}$ (by 5 and 14\,km/s,
respectively, in 2001 and 2003), testifying to a variable rate of
expansion of the layers in which they are formed. Our discovery of
evidence that the wind is variable shows the necessity of spectroscopic
monitoring of this star.

\section{Acknowledgements}

We are grateful to M.V.\,Yushkin for his assistance in the observations
at the 6\,m telescope and in the preliminary data reduction, to
V.E.\,Panchuk for fruitfull discussions of our results,
and to N.S.\,Tavolganskaya for her help with the preparation of this
manuscript. Our study of the spectra of extremely luminous stars are
supported by the Russian
Foundation for Basic Research (project No.\,02--02--16085a), and by the
research program ``Extended Objects in the Universe'' of the
Department of Physical Sciences RAS.

\clearpage
\newpage

\begin{figure}[t]	      		      
\includegraphics[angle=-90,width=1.0\textwidth,bb=37 150 560 800,clip]{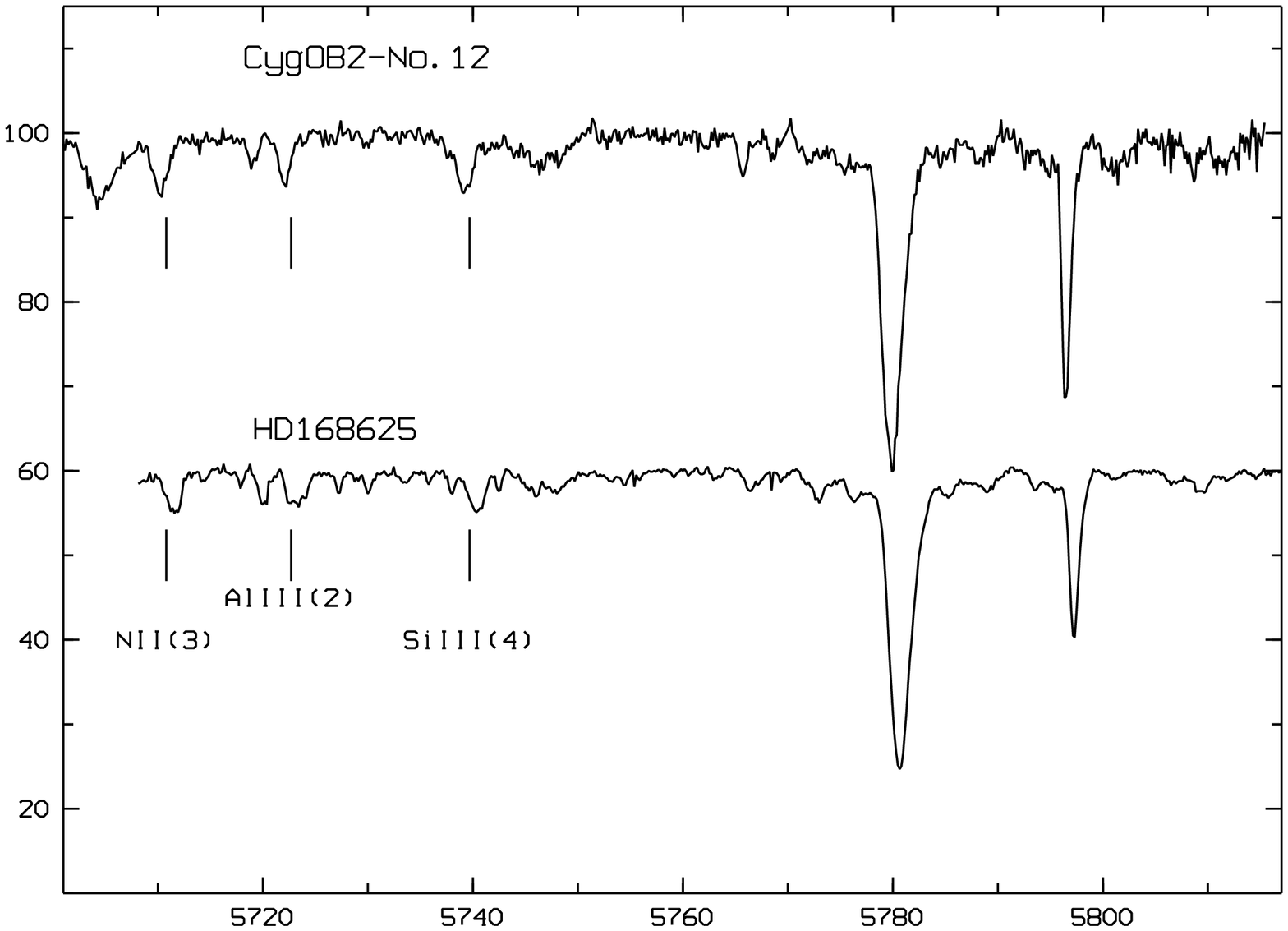}
\caption{Fragment of the spectrum of Cyg\,OB2--No.\,12 taken on June 12,
       2001 (top) compared to the corresponding fragment of the spectrum of the
       $B5Iae^+$ star HD\,168625 (bottom). The stellar lines are marked, with their
       identifications given. The rest of the absorption features are DIBs; the
       deepest lines are at 5780 and 5797\,\AA{} (Table 1).}
\end{figure}

\clearpage
\newpage

\begin{figure}[t]	      		      
\includegraphics[angle=-90,width=1.0\textwidth,bb=37 150 560 800,clip]{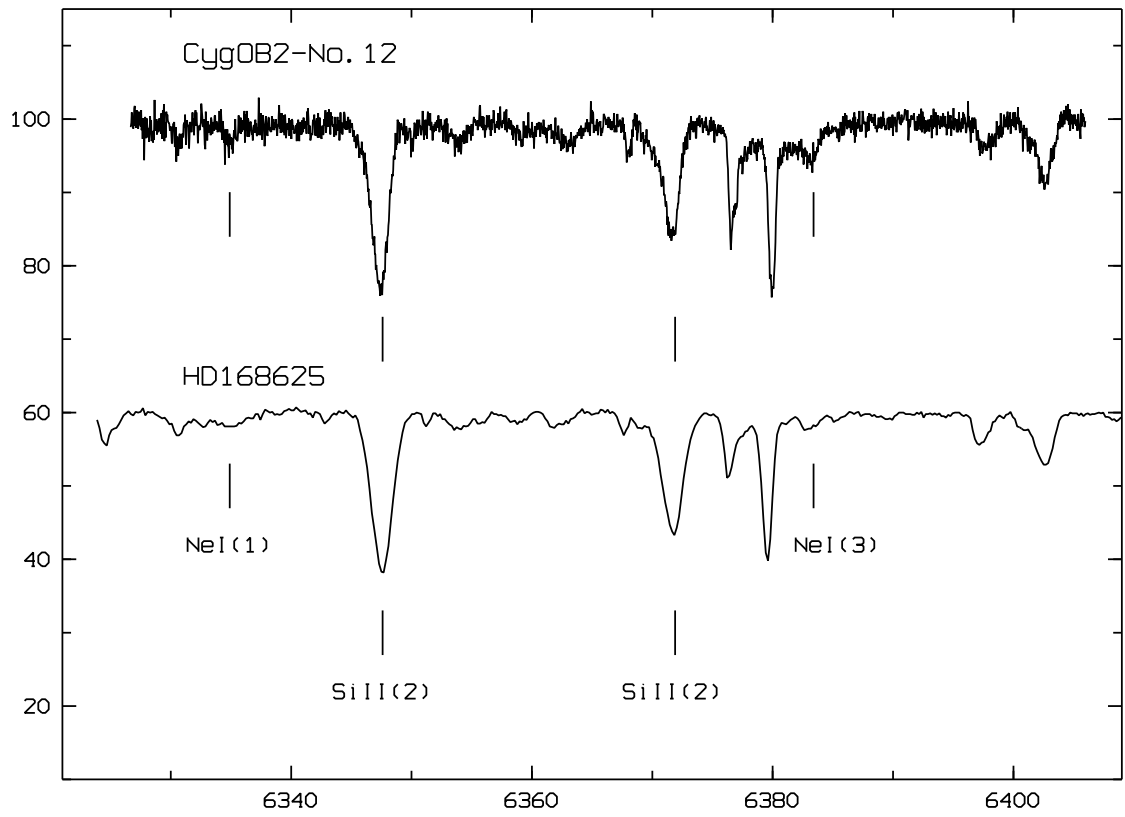}
\caption{Same as Fig.\,1 for the region of the SiII doublet (Table 1).} 
\end{figure}

\clearpage
\newpage

\begin{figure}[t]	      		      
\includegraphics[angle=-90,width=1.0\textwidth,bb=30 150 570 810,clip]{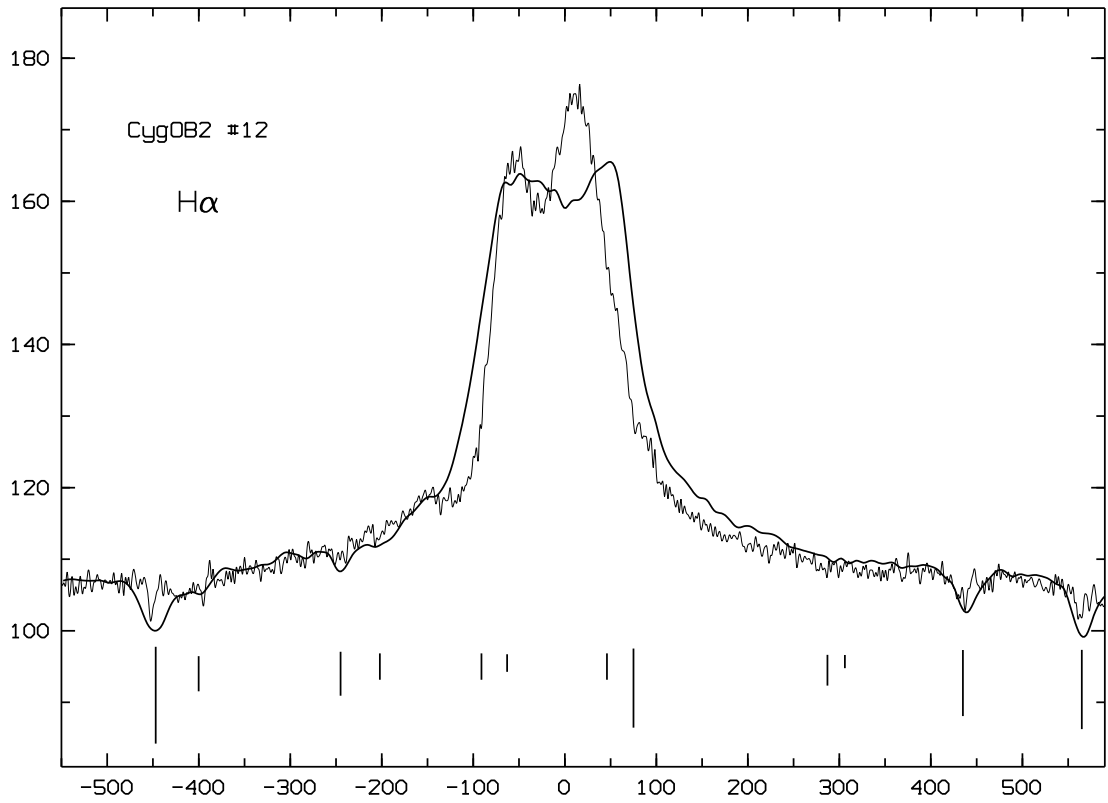}
\caption{H$\alpha$ line profiles ($V_r(r)$) relations) in the spectra of
   Cyg\,OB2--No.\,12. 12 obtained on June 12, 2001 (solid) and April 12, 2003
  (thin). The vertical dashes show the positions of telluric  lines, with
  the dash lengths being proportional to the line depths.}
\end{figure}

\clearpage
\newpage

\begin{figure}[t]	      		      
\includegraphics[angle=0.5,width=1.0\textwidth,bb=37 450 480 810,clip]{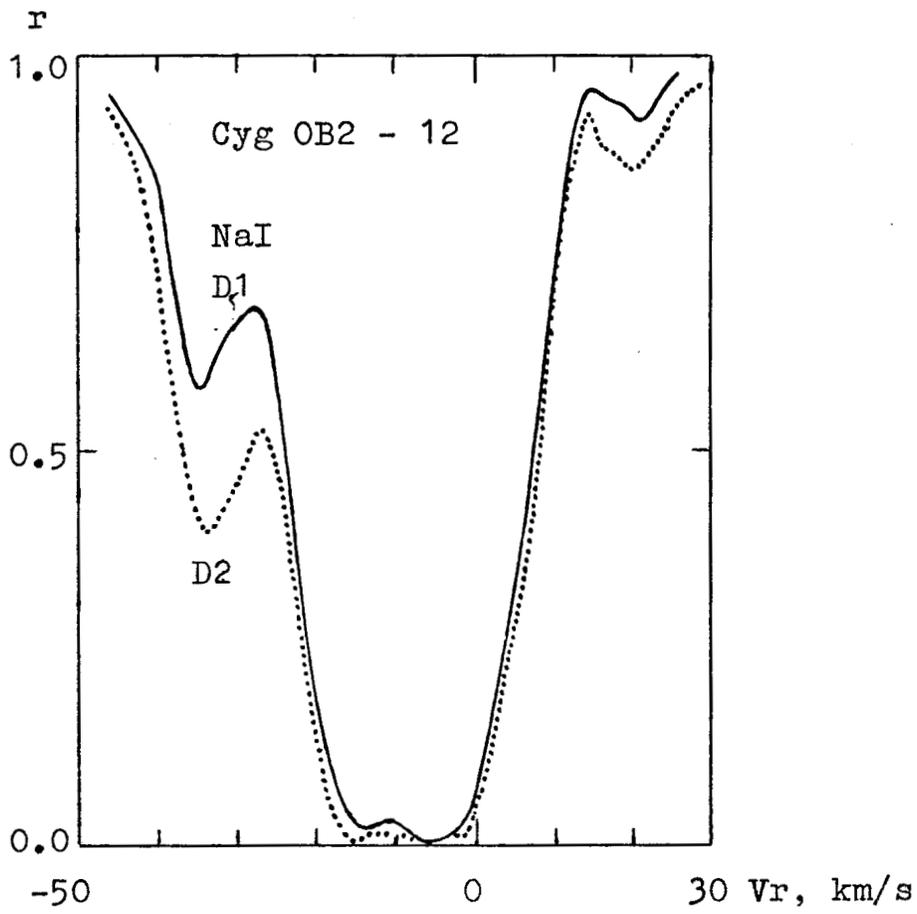}
\caption{Profiles of the D1 (solid) and D2 (dotted) NaI lines
        in the spectrum of Cyg\,OB2--No.\,12 (April 12, 2003).}
\end{figure}

\clearpage
\newpage

\begin{figure}[t]	      		      
\includegraphics[angle=0,width=1.0\textwidth,bb=30 450 560 800,clip]{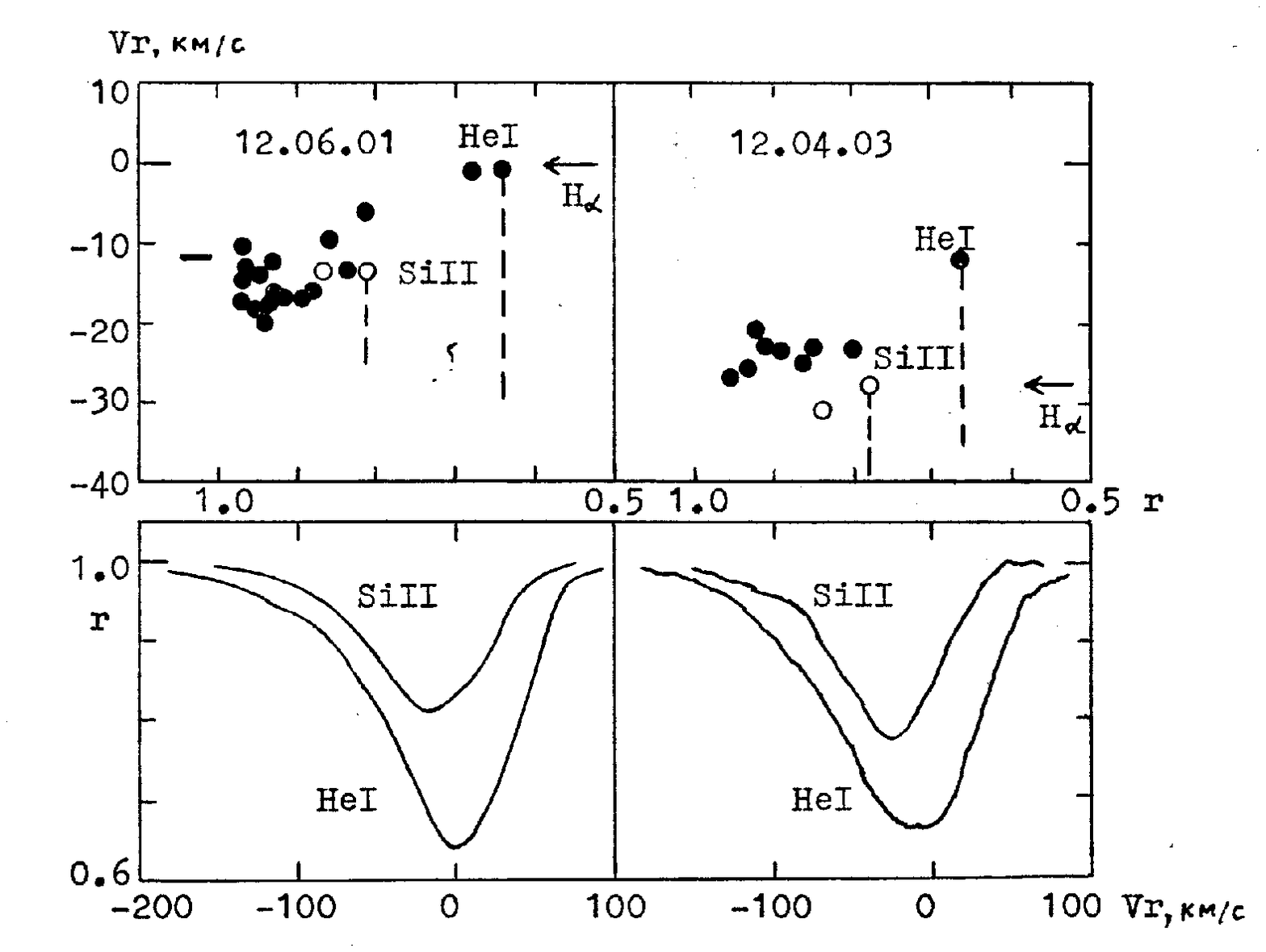}
\caption{Radial velocities in the atmosphere of Cyg\,OB2--No.\,12.
     Top: the heliocentric radial velocities versus the residual intensities.
     Circles correspond to absorption cores, the horizontal dash to the FeII\,7513\,\AA{} 
     emission, and arrows to dips in the peak of the H$\alpha$ profiles. The vertical
     dashed lines show the direction and amount of wing shifts for the
     SiII\,6347\,\AA{} and HeI\,5876\,\AA{} lines, and the horizontal dashed line
     corresponds to  $V_{sys}= -11$\,km/s.
     Bottom: the corresponding profiles of the SiII\,6347\,\AA{} and HeI\,5876\,\AA{} lines.}
\end{figure}

\clearpage
\newpage

\setlongtables
\begin{longtable}{llc@{\quad} @{\quad}l @{\quad}l @{\quad}l@{\quad}@{\quad}l @{\quad}l @{\quad}l@{\quad}}
\caption{Lines indentification, their equivalent widths, $W_{\lambda}$, and depths, $r$, for the Cyg\,OB2--No.12 spectra obtained for 2 dates.
The uncertain values are indicated by colons.}
\\ \hline
\multicolumn{2}{l}{Элемент} & ${\rm \lambda_{lab}}$
      & \multicolumn{3}{c}{\underline{\qquad\qquad 12.06.01 \qquad}}
      & \multicolumn{3}{c}{\underline{\qquad\qquad 12.04.03 \qquad}}\\
\medskip
&  &  & $W_{\lambda}$ & $r$ & $V_r$ &$W_{\lambda}$ & $r$ & $V_r$\\ \hline
1 & 2 & 3 & 4 & 5 & 6 & 7 & 8 & 9 \\ \hline 
\endfirsthead
\hline\multicolumn{9}{l}{\small Table~1, continued} \\ \hline 
 1 & 2 & 3 & 4 & 5 & 6 & 7 & 8 & 9 \\ \hline 
\endhead
\hline\multicolumn{9}{r}{\small to be continued} \\  \hline
\endfoot
\hline
\endlastfoot
   HeI  & (48) &  4921.93 & 0.43: & 0.73  & $-$18:  &&& \\ 
   FeII & (42) &  4923.92 & 0.13: & 0.9:  &       &&& \\ 
   DIB  &      &  4963.90 & 0.17: & 0.86  & $-$12:  &&& \\ 
   DIB  &      &  4984.81 &       & 0.92: &       &&& \\ 
   NII  & (19) &  5001.4: & 0.20: & 0.90: &  $-$5:  &&& \\ 
   NII  & (6)  &  5002.70 &       & 0.93: &	  &&&  \\ 
   NII  & (19,6)& 5005.15 & 0.13: & 0.91  &       &&& \\ 
   NII  & (24) &  5007.33 & 0.08: & 0.95  &	  &&& \\ 
   SII  & (7)  &  5009.56 &       & 0.95: &	  &&&  \\ 
   NII  & (4)  &  5010.62 & 0.12: & 0.94  & $-$18  &&&\\ 
   HeI  & (4)  &  5015.68 & 0.35  & 0.83  & $-$14  &&&\\ 
   FeII & (42) &  5018.44 & 0.12: & 0.91: & $-$ 9: &&&\\ 
   SII  & (7)  &  5032.45 & 0.12: & 0.93: & $-$17: &&&\\ 
   NII  & (4)  &  5045.10 &       & 0.96: &	 &&&   \\ 
   HeI  & (47) &  5047.74 & 0.12: & 0.93: &      &&&\\ 
   FeIII& (5)  &  5073.90 & 0.08: & 0.96: &      &&&\\ 
   FeIII& (5)  &  5086.72 & 0.05: & 0.96: &      &&&\\ 
   FeIII& (5)  &  5127.35 & 0.09: & 0.95: & $-$12: &&&\\ 
   FeIII& (5)  &  5156.12 & 0.09: & 0.92: & $-$10: &&&\\ 
   FeII & (42) &  5169.03 & 0.13: & 0.93  & $-$13  &&&\\ 
   FeIII& (113) & 5235.66 & 0.10: & 0.95: &    &&&\\ 
   FeIII& (113) & 5243.31 &       & 0.98: &    &&&\\ 
   FeIII&      &  5260.34 &       & 0.97  &    &&&\\ 
   FeII & (49,48)& 5316.65: & 0.03: & 0.98: &	   &&&\\ 
   SII  & (38) &  5320.73 & 0.04: & 0.97: & $-$15:  &&&\\ 
   SII  & (38) &  5345.72 & 0.06: & 0.97: & $-$18:  &&&\\ 
   DIB  &      &  5404.50 &       & 0.95: & $-$12:  &&&\\
   DIB  &      &  5418.90 &       & 0.96: &      & 0.07: & 0.92  & $-$16: \\
   SII  & (6)  &  5428.67 & 0.07  & 0.96  & $-$19: & 0.05: & 0.96: & $-$26: \\
   SII  & (6)  &  5432.82 & 0.10  & 0.94  & $-$20: & 0.13  & 0.91  & $-$23  \\
   SII  & (6)  &  5453.83 & 0.24  & 0.88  & $-$16  & 0.21  & 0.86  & $-$25  \\
   SII  & (6)  &  5473.62 & 0.05  & 0.97  & $-$18  & 0.09: & 0.94: &      \\
   DIB  &      &  5487.67 & 0.22  & 0.95  &      &&&\\
   DIB  &      &  5494.10 & 0.07  & 0.92  & $-$14  & 0.08  & 0.90  & $-$13  \\
   NII  & (29) &  5495.67 & 0.06: & 0.97: &      &&& \\
   DIB  &      &  5508.35 &       & 0.95  &       &&&\\
   SII  & (6)  &  5509.72 &       & 0.95: & $-$17:  &&&\\
   DIB  &      &  5512.64 & 0.04  & 0.96  & $-$12: & 0.04  & 0.93  &  $-$8: \\
   CII  & (10) &  5535.35 & 0.06: & 0.97: &       &&&\\
   DIB  &      &  5541.62 &       & 0.97  &       &&&\\
   DIB  &      &  5544.96 & 0.05  & 0.94  & $-$9  & 0.09: & 0.92  & $-$12: \\
   SII  & (6)  &  5564.98 &       &       &      & 0.04: & 0.96: & $-$27: \\
   DIB  &      &  5594.59 &       & 0.98  & $-$12:  &&&\\
   SII  & (11) &  5606.15 & 0.06  & 0.96  & $-$13  & 0.09  & 0.95: & $-$25: \\
   DIB  &      &  5609.73 & 0.04  & 0.98  & $-$10  & 0.04  & 0.96: & $-$10: \\
   SII  & (11) &  5616.64 &       & 0.98: & $-$15: & &&\\
   SII & (14,11)& 5640.1: & 0.17  & 0.89  & $-$10: & 0.23  & 0.89  & $-$24: \\
   SII  & (14) &  5647.03 & 0.06  & 0.96  & $-$18: & 0.07  & 0.93  & $-$26  \\
   SII  & (11) &  5659.99 &       & 0.99: &      & 0.06: & 0.95: & $-$27: \\
   SII  & (11) &  5664.78 &       & 0.98  &      &&&\\
   NII  & (3)  &  5666.63 & 0.13: & 0.92  & $-$17  & 0.11  & 0.92  & $-$27: \\
   NII  & (3)  &  5676.02 & 0.12: & 0.95  & $-$19: & 0.13  & 0.92  & $-$20: \\
   NII  & (3)  &  5679.56 & 0.23: & 0.89  & $-$17  & 0.24  & 0.85  & $-$23  \\
   NII  & (3)  &  5686.21 & 0.13: & 0.96  & $-$22:  &&&\\
   AlIII& (2)  &  5696.60 & 0.10  & 0.93  & $-$13  & 0.12  & 0.91  & $-$23  \\
   DIB  &      &  5705.20 & 0.25  & 0.92  & $-$20: & 0.30  & 0.92  &      \\
   NII  & (3)  &  5710.77 & 0.13  & 0.93  & $-$16  & 0.14  & 0.92  & $-$16  \\
   DIB  &      &  5719.30 & 0.05  & 0.97  & $-$1   &&&\\
   AlIII& (2)  &  5722.73 & 0.09  & 0.94  & $-$20  & 0.06: & 0.96  & $-$15: \\
   SiIII& (4)  &  5739.73 & 0.15  & 0.93  & $-$10  & 0.12  & 0.92  & $-$19: \\
   NII  & (9)  &  5747.30 &       & 0.96  &       &&&\\
   DIB  &      &  5766.16 & 0.05  & 0.95  & $-$8  & 0.06  & 0.96  & $-$13: \\
   DIB  &      &  5769.04 & 0.03  & 0.98  & $-$10: & 0.03  & 0.97: & $-$8: \\
   DIB  &      &  5772.60 & 0.05: & 0.97  & $-$14: & 0.04  & 0.97: & $-$10: \\
   DIB  &      &  5775.78 &       & 0.97: &       &&&\\
   DIB  &      &  5780.37 & 1.03  & 0.60  & $-$9  & 0.95  & 0.62  &  $-$8  \\
   DIB  &      &  5785.05 &       & 0.97: &       &&&\\
   DIB  &      &  5793.22 &       & 0.97: & $-$6:  &&&\\
   DIB  &      &  5795.16 &       & 0.96: & $-$5:  &&&\\
   DIB  &      &  5796.96 & 0.38  & 0.69  & $-$9  & 0.35  & 0.69  &  $-$9  \\
   DIB  &      &  5809.24 & 0.06: & 0.97: &      & 0.05  & 0.96: &  $-$9: \\
   DIB  &      &  5811.96 &       & 0.98  & $-$15:  &&&\\
   DIB  &      &  5818.75 &       & 0.97: & $-$6: & 0.05  & 0.96: & $-$13: \\
   DIB  &      &  5828.46 & 0.06: & 0.97  &       &&&\\
   FeIII& (114)&  5833.93 & 0.05: & 0.97  & $-$15: & 0.07  & 0.96: & $-$15: \\
   DIB  &      &  5842.23 &       & 0.97  &       &&&\\
   DIB  &      &  5844.80 &       & 0.96  & $-$12   &&&\\
   DIB  &      &  5849.80 & 0.16  & 0.88  & $-$12  & 0.14  & 0.86  & $-$12  \\
   HeI  & (11) &  5875.72 & 0.76  & 0.64  & $-$2  & 0.82  & 0.66  & $-$12  \\
   NaI  & (1)  &  5889.95 &       &       &      &       & 0.40  & $-$34  \\
   NaI  & (1)  &  5889.95 & 0.92  & 0.11  & $-$13  & 0.82  & 0.01  &  $-$9  \\
   NaI  & (1)  &  5895.92 &       &       &      &       & 0.58  & $-$34  \\
   NaI  & (1)  &  5895.92 & 0.76  & 0.14  & $-$12  & 0.72  & 0.02  &  $-$9  \\
   DIB  &      &  6005.03 &       & 0.96: & $-$7:  &&&\\
   DIB  &      &  6010.65 & 0.30: & 0.95  & $-$13   &&&\\
   DIB  &      &  6019.36 & 0.03: & 0.97  & $-$14  & 0.03: & 0.98: & $-$13: \\
   DIB  &      &  6027.48 & 0.06: & 0.98  & $-$10  & 0.07  & 0.95  & $-$11  \\
   DIB  &      &  6037.61 & 0.09: & 0.97  & $-$19: & 0.15  & 0.96  &   \\
   OI   & (22) &  6046.4: & 0.03: & 0.99: &       &&&\\
   DIB  &      &  6059.67 & 0.05: & 0.98: & $-$11:  &&&\\
   DIB  &      &  6065.20 &       & 0.98: & $-$18: & 0.04  & 0.97: &  $-$9: \\
   DIB  &      &  6068.20 & 0.02: & 0.99  &       &&&\\
   NeI  & (3)  &  6074.34 & 0.04: & 0.98  & $-$7:  &&&\\
   DIB  &      &  6084.75 & 0.03: & 0.98  &       &&&\\
   PII  & (5)  &  6087.82 & 0.02: & 0.99: &       &&&\\
   DIB  &      &  6089.78 & 0.06  & 0.94  & $-$7  & 0.04  & 0.94  & $-$11  \\
   NeI  & (3)  &  6096.16 & 0.05  & 0.97: & $-$8:  &&&\\
   DIB  &      &  6108.05 &       & 0.98: & $-$10:  &&&\\
   DIB  &      &  6113.20 & 0.05  & 0.96  & $-$14  & 0.06  & 0.96  & $-$13  \\
   DIB  &      &  6116.80 & 0.03: & 0.98  & $-$10:  &&&\\
   DIB  &      &  6118.68:&       & 0.99  &       &&&\\
   DIB  &      &  6139.94 & 0.03  & 0.97  & $-$15  & 0.02  & 0.96  & $-$10: \\
   NeI  & (1)  &  6143.06 & 0.05  & 0.97  & $-$8: & 0.10  & 0.95  & $-$27  \\
   FeII & (74) &  6147.74 &       & 0.99  &        &&&\\
   OI   & (10) &  6156.3: &       & 0.99  &        &&&\\
   OI   & (10) &  6158.18 &       & 0.98  & $-$5:  &&&\\
   DIB  &      &  6161.9: &       & 0.98  & $-$10:  &&&\\
   NeI  & (5)  &  6163.59 &       & 0.98  &        &&&\\
   PII  & (5)  &  6165.59 &       & 0.98: &        &&&\\
   FeIII&      &  6185.26 &       & 0.98: &        &&&\\
   DIB  &      &  6194.73 &       & 0.98  &        &&&\\
   DIB  &      &  6195.96 & 0.12  & 0.86  & $-$12  & 0.12  & 0.80  & $-$12 \\
   DIB  &      &  6203.08 &       & 0.85  & $-$14  & 0.30: & 0.85  & $-$12 \\
   DIB  &      &  6211.66 & 0.03: & 0.97  & $-$14:  &&&\\
   DIB  &      &  6212.90 & 0.02: & 0.98  & $-$9:  &&&\\
   DIB  &      &  6215.79 &       & 0.99  &       &&& \\
   DIB  &      &  6220.81 &       & 0.99  & $-$15:  &&&\\
   DIB  &      &  6223.56 &       & 0.98  &      & 0.04  & 0.95  &  \\
   DIB  &      &  6226.30 &       & 0.99  &       &&& \\
   DIB  &      &  6234.03 &       & 0.95  & $-$14: & 0.05  & 0.95  & $-$12 \\
   DIB  &      &  6236.67 &       & 0.98  &        &&&\\
   FeII & (74) &  6238.39 &       & 0.99: &        &&&\\
   FeII & (74) &  6247.55 &       & 0.99: &        &&&\\
   DIB  &      &  6250.82 &       & 0.98  & $-$12:   &&&\\
   NeI  & (5)  &  6266.50 & 0.04: & 0.97  & $-$14: & 0.05  & 0.97  &   \\
   DIB  &      &  6269.75 & 0.27  & 0.86  & $-$13  & 0.28  & 0.85  & $-$11  \\
   DIB  &      &  6283.85 &       & 0.59  & $-$14  &       & 0.61  & $-$10: \\
   SII  & (26) &  6312.66 &       & 0.98  & $-$16:  &&&\\
   FeII &      &  6317.99 &       & 0.98: &       &&& \\
   DIB  &      &  6324.80 & 0.04: & 0.97: &       &&& \\
   DIB  &      &  6329.97 & 0.03  & 0.98  & $-$11:  &&&\\
   NeI  & (1)  &  6334.43 & 0.03  & 0.98: & $-$9: &       & 0.97: & $-$20: \\
   SiII & (2)  &  6347.10 & 0.39  & 0.81  & $-$14  & 0.40  & 0.78  & $-$28  \\
   DIB  &      &  6353.34 & 0.05  & 0.98  & $-$15: & 0.04  & 0.97  & $-$15: \\
   DIB  &      &  6362.30 & 0.05  & 0.98  & $-$15: & 0.06  & 0.97: & $-$12: \\
   DIB  &      &  6367.25 & 0.04: & 0.97  & $-$13: & 0.04  & 0.95  & $-$8   \\
   SiII & (2)  &  6371.36 & 0.28  & 0.87  & $-$14  & 0.33  & 0.84  & $-$30  \\
   DIB  &      &  6375.95 & 0.11: & 0.93  & $-$10  & 0.13  & 0.86: & $-$14  \\ 
   DIB  &      &  6379.29 & 0.18  & 0.83  & $-$12  & 0.20  & 0.78  & $-$12  \\ 
   NeI  & (3)  &  6382.99 & 0.04  & 0.97  &      & 0.06: & 0.95  & $-$32: \\
   SII  & (19) &  6384.89 &       & 0.99  &	  &&&    \\ 
   DIB  &      &  6397.39 & 0.08  & 0.96  &	   &&&  \\ 
   DIB  &      &  6400.30 &       & 0.98: &	   &&&  \\ 
   NeI  & (1)  &  6402.25 & 0.13  & 0.93  & $-$9    &&&\\ 
   DIB  &      &  6410.18 & 0.04: & 0.98  & 	   &&&  \\ 
   SII  & (19) &  6413.71 & 0.05: & 0.98  & 	   &&&  \\ 
   DIB  &      &  6425.70 & 0.03: & 0.97  & $-$13    &&&  \\ 
   FeII & (40) &  6432.68 & 0.03: & 0.98: & 	   &&&  \\ 
   DIB  &      &  6439.50 & 0.06  & 0.95  & $-$12  & 0.05  & 0.94  & $-$9  \\ 
   DIB  &      &  6445.20 & 0.05  & 0.94  & $-$10  & 0.07  & 0.92  & $-$11  \\ 
   DIB  &      &  6449.14 & 0.04  & 0.96: & $-$11: & 0.05  & 0.96  & $-$10: \\ 
   FeII & (74) &  6456.38 &       & 0.95: & $-$15:   &&& \\  
   NII  & (8)  &  6482.05 &       & 0.95: & $-$13:   &&& \\ 
   NeI  & (3)  &  6506.53 &       & 0.97  & $-$10: &       & 0.93: & $-$26: \\ 
   H$\alpha$ & &  6562.81 & 4.0:  & 1.6   & $-$8  &       & 1.76  & $-$15  \\ 
   CII  & (2)  &  6578.05 & 0.38: & 0.82  & $-$6  & 0.39  & 0.80: & $-$23: \\ 
   CII  & (2)  &  6582.88 & 0.29: & 0.86  & $-$9     &&&\\ 
   DIB  &      &  6597.31 &       & 0.96  & $-$10:    &&&\\ 
   NeI  & (6)  &  6598.95 &       & 0.98  & 	    &&&\\ 
   NII  & (31) &  6610.57 &       & 0.99: &	    &&&\\ 
   DIB  &      &  6613.56 & 0.42  & 0.70  & $-$9  & 0.39  & 0.68  & $-$12 \\ 
   DIB  &      &  6632.85 &       & 0.99: & 	     &&&\\ 
   OII  & (4)  &  6640.90 &       & 0.99: & 	     &&&\\ 
   DIB  &      &  6646.03 &       & 0.99: & 	     &&&\\ 
   DIB  &      &  6660.64 & 0.06  & 0.93  & $-$11  & 0.07  & 0.92  & $-$10 \\ 
   DIB  &      &  6665.15 &       & 0.98  & 	     &&&\\ 
   DIB  &      &  6672.15 & 0.05  & 0.96  & $-$13  &&&\\ 
   HeI  & (46) &  6678.15 & 0.75  & 0.68  & $-$2  &&&\\ 
   DIB  &      &  6689.30 & 0.03: & 0.99  & $-$13: &&&\\ 
   DIB  &      &  6694.48 & 0.02: & 0.99  & $-$17: &&&\\ 
   DIB  &      &  6699.26 & 0.07  & 0.95  & $-$11  &&&\\ 
   DIB  &      &  6701.98 & 0.02  & 0.98  & $-$15  &&&\\ 
   DIB  &      &  6709.39 & 0.02: & 0.98  & $-$6: &&&\\ 
   DIB  &      &  6729.28 & 0.02: & 0.98  & $-$11: &&&\\ 
   DIB  &      &  6737.13 & 0.02: & 0.99  & $-$13: &&&\\ 
   DIB  &      &  6740.99 & 0.03: & 0.98  &      &&&\\ 
   DIB  &      &  6767.74 & 0.01: & 0.99  & $-$17: &&&\\ 
   DIB  &      &  6770.05 & 0.03: & 0.98  & $-$9  &&&\\ 
   DIB  &      &  6788.66 & 0.02: & 0.99  & $-$7: &&&\\ 
   DIB  &      &  6792.52 & 0.03  & 0.98  & $-$12  &&&\\ 
   DIB  &      &  6795.24 & 0.03  & 0.98  & $-$6: &&&\\ 
   DIB  &      &  6801.37 & 0.02: & 0.98  & $-$9  &&&\\ 
   DIB  &      &  6810.5: &       & 0.98: &      &&&\\
   DIB  &      &  6827.30 & 0.02: & 0.98  & $-$14  &&&\\ 
   DIB  &      &  6843.60 & 0.06  & 0.96  & $-$19  &&&\\ 
   DIB  &      &  6852.67 & 0.02  & 0.98  &	 &&&\\ 
   DIB  &      &  6860.02 & 0.05  & 0.97  & $-$17  &&&\\ 
   DIB  &      &  6862.53 & 0.03: & 0.99  &      &&&\\ 
   HeI  & (10) &  7065.32 & 0.36: & 0.82  & $+$2: &&&\\ 
   DIB  &      &  7357.60 &       & 0.91  & $-$14: &&&\\ 
   NI   & (3)  &  7468.31 & 0.05  & 0.98  & $-$15: &&&\\ 
   DIB  &      &  7494.89 &       & 0.98: & $-$9: &&&\\ 
   FeII &      &  7495.63 &       & 1.01  &      &&&\\ 
   FeII &      &  7513.17 &       & 1.03  & $-$12  &&&\\ 
   DIB  &      &  7559.35 &       & 0.97  & $-$12  &&&\\ 
   DIB  &      &  7562.3: & 0.25: & 0.94  & $-$13: &&&\\ 
   DIB  &      &  7581.30 & 0.08: & 0.97  & $-$12: &&&\\ 
   KI   & (1)  &  7664.91 & 0.70: & 0.4:  & $-$9: &&&\\ 
   DIB  &      &  7721.85 & 0.07  & 0.96  & $-$10  &&&\\ 
   OI   & (1)  &  7771.94 & 1.23$^1$ & 0.76 & $+$3:&&& \\ 
   OI   & (1)  &  7774.2: &       & 0.75  & 	  &&& \\ 
   DIB  &      &  7832.81 & 0.05  & 0.97  & $-$10  &&&\\ 
\hline
\multicolumn{9}{l}{\small $^1$  -- the sum of $W_{\lambda}$  of the OI--triplet lines at 7773\,\AA{}}
\end{longtable}	       			     
\label{lines}

\clearpage
\newpage

\begin{table}[hbtp]
\caption{Heliocentric radial velocities, $V_r$, for individual lines and
        groups of lines in the spectrum of Cyg\,OB2--No.\,12 (uncertain values indicated by colons)}
\begin{tabular}{lll}
\hline \rule{0pt}{5pt}
Lines &  \multicolumn{2}{c}{${\rm V_r}$\,, km/s} \\
\cline{2-3}
      &  12.06.01    &  12.04.03 \\
\hline
\multicolumn{3}{l}{\underline{Emissions}} \\
${\rm H_{\alpha}}$      & $-$10  &  $-$15 \\
 FeII                &  $-$12: &   \\
\multicolumn{3}{l}{\underline{Photospheric absorptions}} \\
NII, OII, SII и др.     & $-$16  & $-$25 \\
SiII\,(2)  &$-$14 &$-$29 \\
CII\,(2)   &$-$11 &  $-$24:   \\
HeI\,5876  & $-$1 & $-$12  \\
${\rm H_{\alpha}}$ & $-$140:,  0: &  $-$125:, $-$28 \\  
\multicolumn{3}{l}{\underline{IS absorptions}} \\
NaI\,(1) & $-$11 & $-$34, $-$9, 20, \\
KI\,(1)  &  $-$10:  & \\
DIB      &  $-$10 & $-$11 \\
\hline
\end{tabular}
\end{table}

\end{document}